\documentclass{article}
\usepackage{ijcai22}

\usepackage{times}
\usepackage{soul}
\usepackage{url}
\usepackage[hidelinks]{hyperref}
\usepackage[utf8]{inputenc}
\usepackage[small]{caption}
\usepackage{graphicx}
\usepackage{amsmath}
\usepackage{amsthm}
\usepackage{booktabs}
\usepackage{algorithm}
\usepackage{algorithmic}
\usepackage{makecell}
\usepackage{float}
\usepackage{color}
\usepackage{amssymb}
\usepackage{multirow}
\title{Self-supervised Domain Adaptation for Breaking the Limits of Low-quality Fundus Image Quality Enhancement}

\author{
Qingshan Hou$^{1,2}$
\and
Peng Cao$^{1,2}$\footnote{Corresponding author}\and
Jiaqi Wang$^{1,2}$\and
Xiaoli Liu$^{3}$\and
Jinzhu Yang$^{1,2}$\And
Osmar R. Zaiane$^{4}$
\affiliations
$^1$Computer Science and Engineering, Northeastern University, Shenyang, China\\
$^2$Key Laboratory of Intelligent Computing in Medical Image of Ministry of Education, Northeastern University, Shenyang, China\\
$^3$DAMO Academy, Alibaba Group, China\\
$^4$Alberta Machine Intelligence Institute, University of Alberta, Edmonton, Alberta, Canada
\emails
houqingshancv@gmail.com,
caopeng@mail.neu.edu.cn,
wjq010222@gmail.com,
liuxiaoli.lxl@alibaba-inc.com
yangjinzhu@cse.neu.edu.cn,
zaiane@cs.ualberta.ca
}
\begin{document}

\maketitle

\begin{abstract}
Retinal fundus images have been applied for the diagnosis and screening of eye diseases, such as Diabetic Retinopathy (DR) or Diabetic Macular Edema (DME). However, both low-quality fundus images and style inconsistency potentially increase uncertainty in the diagnosis of fundus disease and even lead to misdiagnosis by ophthalmologists. Most of the existing image enhancement methods mainly focus on improving the image quality by leveraging the guidance of high-quality images, which is difficult to be collected in medical applications. In this paper, we tackle image quality enhancement in a fully unsupervised setting, i.e., neither paired images nor high-quality images. To this end, we explore the potential of the self-supervised task for improving the quality of fundus images without the requirement of high-quality reference images. Specifically, we construct multiple patch-wise domains via an auxiliary pre-trained quality assessment network and a style clustering. To achieve robust low-quality image enhancement and address style inconsistency, we formulate two self-supervised domain adaptation tasks to disentangle the features of image content, low-quality factor and style information by exploring intrinsic supervision signals within the low-quality images. Extensive experiments are conducted on EyeQ and Messidor datasets, and results show that our DASQE method achieves new state-of-the-art performance when only low-quality images are available. 
\end{abstract}

\section{Introduction}
Medical fundus images have been extensively used for clinical analysis of various ocular diseases \cite{teo2021global,dai2021deep,li2019canet}. However, the real clinical fundus datasets usually contain a large number of low-quality images. The quality of fundus images is critical to the diagnosis and screening of eye diseases. In contrast, the low-quality fundus images easily mislead the clinical diagnosis and lead to unsatisfactory results of downstream tasks like blood vessel segmentation. Existing deep learning methods \cite{shen2020modeling,cheng2021secret,gui2021review} rely on a large amount of high-quality fundus images or paired images, which limits their practicality and generalization in clinical applications due to lack of availability of high-quality fundus images. To this end, this paper explores a new perspective: could a model achieve quality enhancement without requiring high-quality images? 

\begin{figure}[t]
\centerline{\includegraphics[width=0.9\columnwidth]{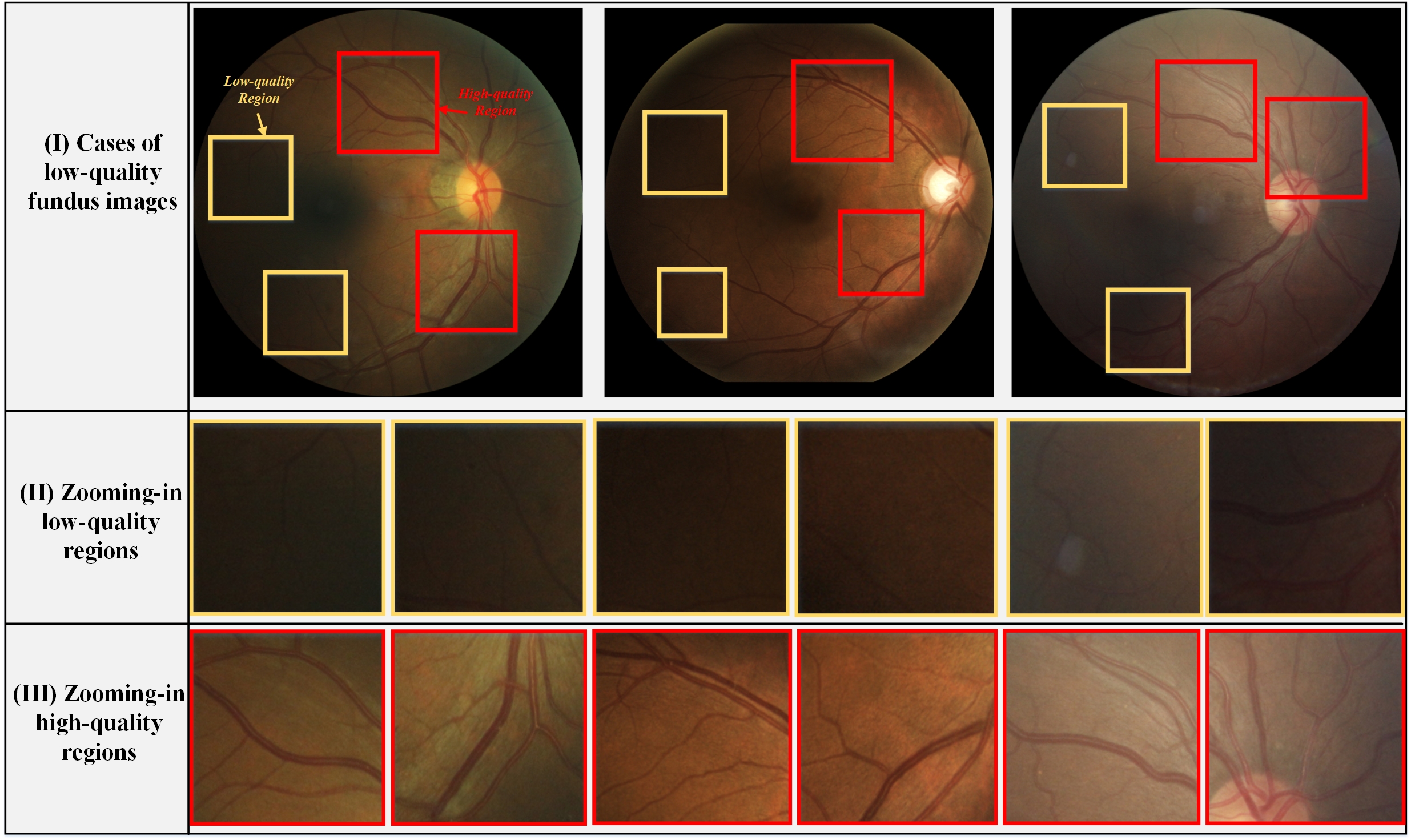}}
\caption{Some cases of low-quality fundus images, where the red box and the yellow box mark the different quality regions in the low-quality images, respectively. The main low-quality factors include uneven illumination, noticeable blur and artifacts.}
\label{fig1}
\end{figure}

\begin{figure*}[t]
\centerline{\includegraphics[width=2\columnwidth]{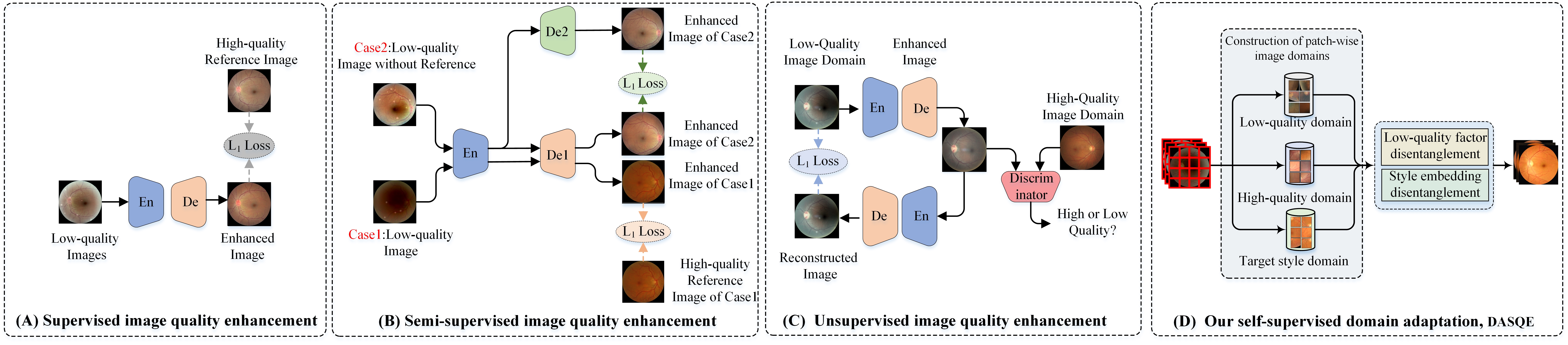}}
\caption{Comparison of different image quality enhancement schemes. Previous fundus image quality enhancement methods (A) require paired images of high-quality and low-quality as training data, which are often difficult to acquire. Although these semi-supervised and unsupervised methods (B) and (C) eliminate the need for paired images, they still need the guidance of high-quality images. (D) In contrast, our method is a truly unsupervised image quality enhancement, requiring neither paired fundus images nor high-quality images.} 
\label{fig2}
\end{figure*}

In low-quality fundus images, we observe some interesting phenomena shown in Fig.\ref{fig1}: not all the regions are low quality, as shown in Fig.\ref{fig1}(I). Besides, there are significant differences among regions in low-quality images with respect to the image styles, as shown in Fig.\ref{fig1}(III). Taking all of the above into consideration, the major challenges of the fundus image quality enhancement lie in: How to disentangle the low-quality factors and image content information for simultaneously 1) improving the quality and 2) unifying the style of images under the condition that only low-quality images are available. Specifically, inter-domain variations include two aspects: the variation between high-quality domain and low-quality domain, and the variation between source high-quality style domain and target high-quality style domain.

To the best of our knowledge, this is the first work to rethink the image quality enhancement problem from a self-supervised learning perspective, without the requirement of any high-quality image. Based on the design of the self-supervision, we propose two assumptions to formulate the self-supervised image quality enhancement: 1) the patches from the same position of the source and translated images should have consistent content. Hence, the low-quality image patches are assumed to be composed of low-quality factors and image content embedding. The procedure of image quality enhancement can be formulated as disentangling the quality factors from the  image content. 2) The low-quality image patches consist of style embedding and image content embedding. To unify the styles, we aim to factorize style-related and content-related embeddings without any auxiliary supervisory signals. Specifically, we propose a high-quality unaware self-supervised domain adaptation framework, named DASQE. The proposed framework first detects the high-quality regions inside the low-quality images by a pre-trained quality assessment network. Given the patch-wise high-quality domain consisting of high-quality patch-wise images, we obtain multiple style domains by clustering and choose a target style domain consisting of high-quality patch-wise images with uniform illumination style. Then, we disentangle patch-wise images into content-related, quality-related and target style-related embedding by separate encoders, and further reconstruct them into original images by different image generators to reduce quality and style variations. The overall procedure involves the cycle consistency loss and the adversarial losses in the latent and image spaces to achieve a self-supervised quality enhancement without any explicit supervision of high-quality data.

In summary, our contributions can be summarized as follows. 
1) A major limitation of most current low-quality fundus image enhancement methods is that they rely on the guidance of paired images or the presence of high-quality images. Hence, we propose a novel medical image quality enhancement method to loosen the requirement of pair-wise training images and only require low-quality input images. To the best of our knowledge, this is the first attempt to apply a self-supervised reference-free method coupled with domain adaptation to the quality enhancement task. We tackle image quality enhancement in a fully unsupervised setting, i.e., neither paired images nor high-quality images are needed. It is a truly unsupervised fundus image quality enhancement. 2) To enhance low-quality fundus images while preserving pathological features and major retinal structures, we propose a  representation decoupling strategy for disentangling content-related, style-related and quality-related embeddings. The low-quality fundus images are enhanced by recombining the learned content-related and target style-related embeddings. It is worth mentioning that our method not only improves the quality of fundus images by the proposed strategy, but also unifies the image style to eliminate style variations. 3) Our self-supervised version without the guidance of any high-quality data significantly outperforms the state-of-the-art methods by a considerable 28.47 PSNR / 91.4\% SSIM and 29.36 PSNR / 89.7\% SSIM on Messidor and EyeQ benchmark datasets, a 1.11/2.63 PSNR and 0.6/1.4\% SSIM improvement compared to the previous best results. Moreover, the proposed method is easily expanded to the quality enhancement task of other medical images. We also further verify that our method is beneficial for a variety of fundus imaging analysis tasks, such as retinal vessel segmentation, lesion segmentation and disease grading.  

\begin{figure*}[t]
\centerline{\includegraphics[width=1.8\columnwidth]{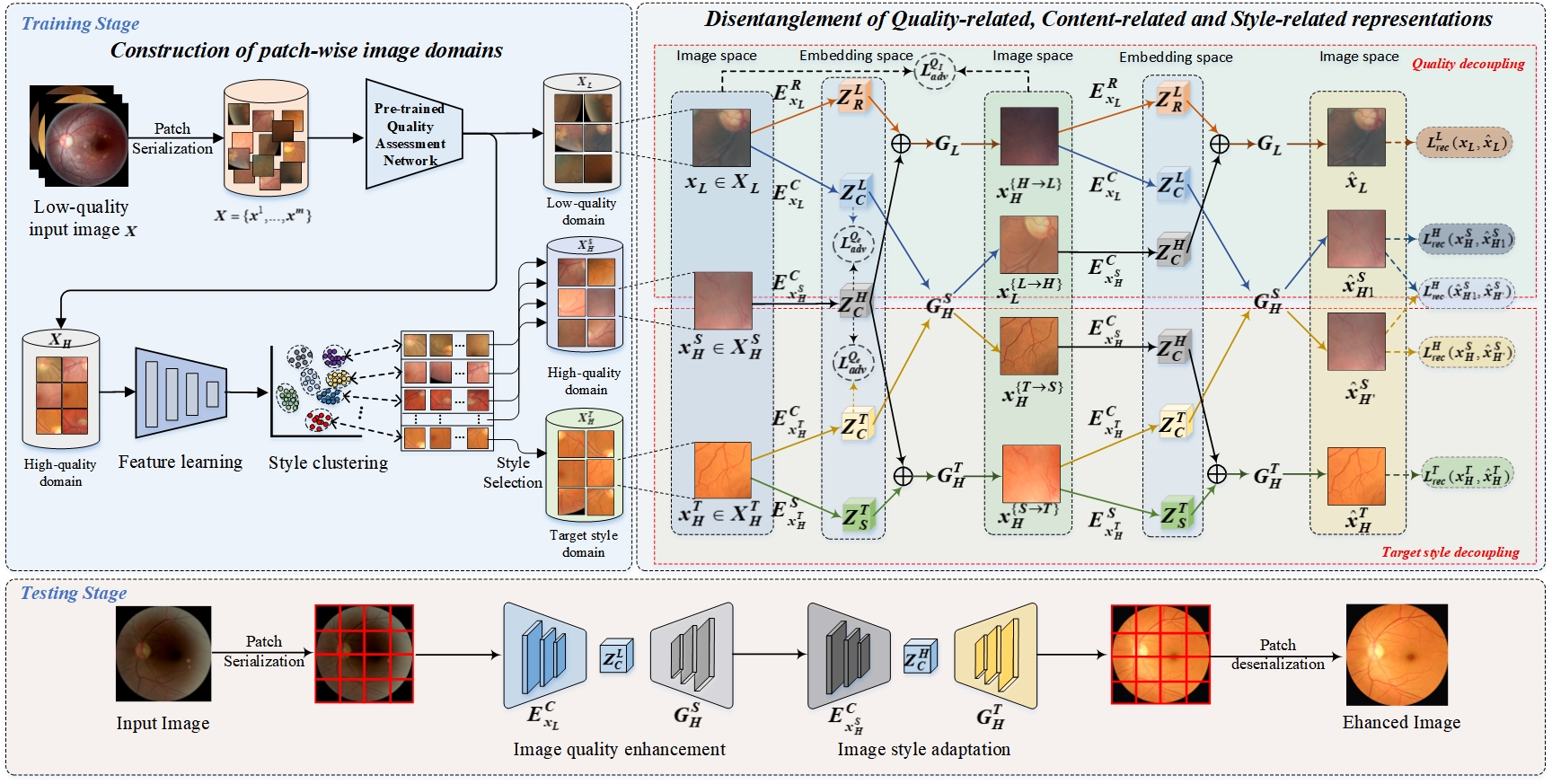}}
\caption{The overall architecture of the proposed framework. It involves two stages: \textbf{Construction of patch-wise image domains:}The low-quality input images $\boldsymbol{X}$ are first serialized as the patch-wise fundus images $\boldsymbol{X}=\left\{x^1, \ldots, x^m\right\}$. With the help of the pre-trained quality assessment network, the patch-wise fundus images are divided into low-quality and high-quality domains, e.g. $\boldsymbol{X_L},\boldsymbol{X_H}$. A style clustering is performed on $\boldsymbol{X_H}$, and a target style domain $\boldsymbol{X^T_H}$ is determined. \textbf{Decoupling of multiple features:} We factorize content-related, quality-related and style-related for patch-wise domains through the supervisions derived from the image data itself to enhance the image quality and align style features. The framework involves content-related encoders \{$\boldsymbol{E}^C_{x_L}$, $\boldsymbol{E}^C_{x^S_H}$,$\boldsymbol{E}^C_{x^T_H}$\} for extracting latent content embedding, quality-related encoder $\boldsymbol{E}^R_{x_L}$ for extracting low-quality factor embedding, target style-related encoder $\boldsymbol{E}^S_{x^T_H}$ for extracting target style embedding, low-/high-quality image generators $\boldsymbol{G}_L$ and $\boldsymbol{G}^S_H$ for generating different quality images, and high-quality target style image generator high-quality style images.}
\label{overview}
\end{figure*}

\section{Related Work}
The quality enhancement methods for fundus images mainly contain traditional non-parametric methods based on hand-crafted priors and data-driven methods based on deep learning. For example, \cite{shome2011enhancement} enhance the contrast of fundus images by the contrast limited adaptive histogram equalization method. \cite{tian2017global} design a global and local contrast enhancement method for the quality enhancement of non-uniform illumination images. \cite{zhou2017color} adjust the luminosity of fundus images based on the luminance gain matrix to achieve enhancement of fundus images with uneven illumination. However, these methods rely heavily on hand-crafted priors, which are hardly applicable to all cases of low-quality fundus image enhancement, such as artifacts. In addition, they also do not provide any learnable parameters to prevent the over-enhancement of fundus images. Recently, deep learning has shown its advantages in a wide range of fields, such as vessel segmentation \cite{jin2019dunet}, lesion detection \cite{wang2018weakly}, disease classification \cite{cao2022collaborative}. For the quality enhancement of fundus images, the most common quality enhancement methods \cite{wang2021joint,shen2020modeling} based on deep learning belong to the fully supervised learning scheme (Fig.\ref{fig2}(A)), which requires supervision with high-quality reference images corresponding to the input during the training stage. However, obtaining paired fundus images is expensive and time-consuming, resulting in limited applications due to the requirement for pairwise training data. In addition, the supervised enhancement methods limit their generalization and practicality across different datasets owing to inter-domain variations between datasets. To loosen the limitations of paired fundus images, a semi-supervised quality enhancement method (Fig.\ref{fig2}(B)) \cite{cheng2021secret} is proposed, which requires only a portion of paired images during the training phase. Inspired by generative adversarial learning, some unsupervised methods (Fig.\ref{fig2}(C)) \cite{you2019fundus,ma2021structure} based on bi-directional GAN have been proposed to enhance low-quality fundus images, which eliminates the requirement for paired fundus images. However, such methods usually require learning knowledge representation in the high-quality domain, and transferring it to the low-quality domain. Its heavy reliance on high-quality data proves to be the pain point of image quality enhancement methods. 

Furthermore, most existing unsupervised methods are usually under-constrained, which may introduce undesirable artifacts or fail to preserve the fine retinal structures and pathological signatures in the real clinical fundus images, and the high-quality image domains are similarly uncommon. From the above analysis and comparison of related work, further exploration of unsupervised quality enhancement method without any guidance of high-quality images is significant and needed for adapting the extensive clinical applications. 

\section{Methodology}
Our main goal is to develop a clinically oriented fundus image quality enhancement method, and uniform illumination style of fundus images. In this section, we first describe the scheme for the construction of the patch-wise domains. Subsequently, we introduce the specific details of multiple feature decoupling, mainly including quality feature decoupling and target style feature decoupling. The quality feature decoupling is designed to remove low-quality interference indicators from patch-wise low-quality images, whereas the style decoupling aims to eliminate style variations among patch-wise images. Finally, we provide the optimization process of the proposed method, the overview of which is illustrated in Fig.\ref{overview}.

\subsection{Construction of patch-wise image domains}
The goal of this part is to obtain the patch-wise low-/high-quality domain and select a high-quality target style domain which is beneficial to the ophthalmologist's diagnosis. Formally, the low-quality fundus images $\boldsymbol{X}$ are serialized into a patch-wise image set $\boldsymbol{X}=\left\{x^1,\ldots, x^m\right\}$ as shown in Fig.\ref{overview}. Then, $\boldsymbol{X}$ are fed into a pre-trained quality assessment network to obtain the patch-wise low-quality image domain $\boldsymbol{X}_L = \left\{x^1_L,\ldots, x^n_L\right\}$ and high-quality image domain $\boldsymbol{X}_H = \left\{x^1_H,\ldots, x^{n'}_H\right\}$, where $m=n+n'$. For quality assessment network, a base DesNet121 model is trained based on the EyeQ dataset \cite{fu2019evaluation} with high-/low-quality binary labels. For the patch-wise high-quality domains $\boldsymbol{X}_H$, there are several distinct illumination styles. This poses significant challenges to the enhancement of low-quality images. In order to align the style of patch-wise images, it is essential to construct a target style domain for the subsequent style decoupling. More specifically, our primary goal is to learn an embedding function De($\cdot$) mapping $\boldsymbol{X}_H$ to embeddings $\boldsymbol{Z}=\left\{z_1,z_2,...,z_j\right\}$ in a \emph{D}-dimension representation space. The loss $\mathcal{L}_{e}$ of latent embedding learning is defined as:

\begin{equation}
\mathcal{L}_{e}=\left\|{\text{De}}\left({\text{En}}({x^i_H})\right)-{x^i_H}\right\|_2^2,
\end{equation}

\noindent where $x^i_H$ indicates the patch-wise images from $\boldsymbol{X}_H$, De($\cdot$) and En($\cdot$) denote the encoder and decoder, respectively. Subsequently, we cluster the obtained embeddings $\boldsymbol{Z} = \text{En}(\boldsymbol{X}_H)$ to obtain the style domains $\left\{\boldsymbol{X}_{S1}, \boldsymbol{X}_{S2}, \cdots, \boldsymbol{X}_{Sn}\right\}$. Finally, we select a style domain containing the highest number of patches as the target style domain $\boldsymbol{X}^T_H$, and the others are considered as source style domains $\boldsymbol{X}^S_H$.

\subsection{Multiple Feature Decoupling}
To maintain the consistency of image-content and enforce the disentanglement of the low-quality factors and the target style during quality enhancement without any extra annotations, the multiple feature decoupling consists of two main aspects: 1) the purpose of the the quality decoupling is to eliminate interference factors from patch-wise low-quality images, and 2) as shown in Fig\ref{fig1} (III), we consider that the original patch style may be changed during the quality decoupling, and in light of the inconsistency of the high-quality patch styles obtained from low-quality images. The target style decoupling is introduced for aligning the style of patch-wise images.

We are given the unpaired high-quality patch-wise image $x^S_H \in \boldsymbol{X}^S_H$, low-quality patch-wise image $x_L \in \boldsymbol{X}_L$, and target style patch-wise image $x^T_H \in \boldsymbol{X}^T_H$. The encoders $E^C_{x_L}$, $E^C_{x^S_H}$, and $E_{x^T_H}^C$ are employed to extract the content embedding $Z^L_C$, $Z^H_C$, and $Z^T_C$ of $x_L$, $x^S_H$, and $x^T_H$, respectively. For $x_L$ and $x^T_H$, quality-related encoder $E^R_{x_L}$ and style-related encoder $E^S_{x^T_H}$ are designed to extract low-quality factor embedding $Z^L_R$ and style-related embedding $Z^T_S$. For the quality decoupling, the process of feature extraction and image generation can be formulated as:
\begin{equation}
\begin{aligned}
& x_L^{\{L \rightarrow H\}}=G_H^S\left(E_{x_L}^C\left(x_L\right)\right) \\
& x_H^{\{H \rightarrow L\}}=G_L\left(E_{x_L}^R\left(x_L\right) \oplus E_{x_H^S}^C\left(x_H^S\right)\right),
\end{aligned}
\label{eq2}
\end{equation}
\noindent where $\oplus$ indicates the channel-wise concatenation. The learned $Z^H_C$ and $Z^L_R$ are recombined to generate low-quality image $x^{\left\{H\rightarrow L\right\}}_H$ via generator $G_L$. For high-quality images $x^{\left\{L\rightarrow H\right\}}_L$, $G^S_H$ only utilizes the content embedding $Z^L_C$.

After decoupling the low-quality factors, there is a visible perceptual disparity among the high-quality patches, which suggests the presence of a style gap among the image patches. Hence, the target style decoupling is proposed for aligning image style. Analogous to the quality decoupling, generator $G_H^S$ is used to transform a high-quality image $x^T_H$ with the target style into a general high-quality image $x^{\{T \rightarrow S\}}_H$, while generator $G^T_H$ jointly takes the $Z^T_S=E_{x^T_H}^S\left(x^T_H\right)$ and $Z^H_C=E_{x_H^S}^C\left(x_H^S\right)$ to generate a new style-changing image $x^{\{S \rightarrow T\}}_H$. We align different style domains by extracting the style embedding of target style domain, and re-feeding the content embeddings from source domains with the decoupled target style embedding to be reconstructed into the image space, thereby alleviating the domain shift problem.

Meanwhile, we define feature-level adversarial loss $\mathcal{L}^{Q_e}_{adv}$ to further encourage the content-related encoders $E^C_{x_L}$, $E^C_{x^S_H}$ and $E^C_{x^T_H}$ to pull their embedding $Z^L_C$, $Z^H_C$ and $Z^T_C$ together, so that inter-domain content variations are reduced.
 
\begin{equation}
\begin{aligned}
\mathcal{L}_{adv}^{{Q_e}} & =\mathbb{E}_{x_L}\left[\frac{1}{2} \log {D}_{{C}}\left({Z}_{{C}}^L\right)+\frac{1}{2} \log \left(1-{D}_{{C}}\left({Z}_{{C}}^L\right)\right)\right] \\
& +\mathbb{E}_{x_H^S}\left[\frac{1}{2} \log {D}_{{C}}\left({Z}_C^H\right)+\frac{1}{2} \log \left(1-{D}_C\left({Z}_{{C}}^{{H}}\right)\right)\right]\\
& +\mathbb{E}_{x^T_H}\left[\frac{1}{2} \log {D}_{{C}}\left({Z}_{{C}}^T\right)+\frac{1}{2} \log \left(1-{D}_{{C}}\left({Z}_{{C}}^T\right)\right)\right],
\end{aligned}
\end{equation}

\noindent where $D_C$ denotes a content representation discriminator. 

In addition to the adversarial loss on the embedding, we define an image-level adversarial loss $\mathcal{L}^{Q_I}_{adv}$ to constrain the generation of $x_L^{\{L \rightarrow H\}}$, $x_H^{\{H \rightarrow L\}}$ $x_H^{\{T \rightarrow S\}}$ and $x_H^{\{S \rightarrow T\}}$, which can force the content-related, quality-related and target style-related encoders to capture their respective embedding.

\begin{equation}
\begin{aligned}
\mathcal{L}_{adv}^{Q_I}&=\mathbb{E}_{x^S_H}\left[\log D_I(x^S_H)\right]+\mathbb{E}_{x^L/x^T_H}\left[\log D_I(x^L/x^T_H)\right]\\
&+\mathbb{E}_{\{x^S_H, x_L/x^T_H\}}\left[\log \left(1-D_{I}\left(x_H^{\{H \rightarrow L\}}/x_H^{\{S \rightarrow T\}}\right)\right)\right]\\
&+\mathbb{E}_{\{x_L/x^T_H,x^S_H\}}\left[\log \left(1-D_{I}\left(x_L^{\{L \rightarrow H\}}/x_L^{\{T \rightarrow S\}}\right)\right)\right],
\end{aligned}
\end{equation}

\noindent where $D_I$ denotes an image-level domain discriminator. 

Besides the adversarial loss for constraining the image generation in the image space, to supervise the adversarial image generation and fully exploit the low-quality images for the quality enhancement model, self-supervised learning can be naturally harnessed for providing additional supervision. A re-feeding strategy is designed to re-disentangle and reconstruct new generated images for image-level self-supervision in the original image space and the disentangled feature space. Specifically, we reconstruct $x_L^{\{L \rightarrow H\}}$ and $x_H^{\{H \rightarrow L\}}$ as $\hat{x}_L$ and $\hat{x}^S_{H1}$ in the original image space via the encoders \{$E_{x_L}^C$, $E_{x_L}^R$, $E_{x_H^S}^C$\} as well as the generators \{$G_L$, $G_H^S$\} according to the same procedure as Eq.\ref{eq2}. Hence, the quality-related self-supervision loss $\mathcal{L}_Q$ for supervision on the reconstructed images can be defined as:
\begin{equation}
\begin{aligned}
\mathcal{L}_Q&=\mathcal{L}^L_{rec}\left(x_L,\hat{x}_L\right)+\mathcal{L}^H_{rec}\left(x^S_H,\hat{x}^S_{H1}\right)\\
&=\mathbb{E}_{x_L}\left[\|\hat{x}_L-x_L\|_1\right]+\mathbb{E}_{x^S_{H}}\left[\|\hat{x}^S_{H1}-x^S_H\|_1\right],
\end{aligned}
\end{equation}
\noindent where $\|\cdot\|_1$ indicates the L$_1$ norm that is widely adopted in self-supervised learning for preserving the consistency of the reconstructed images. Similarly, due to the lack of reference images corresponding to the inputs, we reconstruct $x^{\{T \rightarrow S\}}_H$ and $x^{\{S \rightarrow T\}}_H$ into the space of the original images $\hat{x}^T_H$ and $\hat{x}^S_{H'}$. An identical style-related self-supervised learning loss $\mathcal{L}_T$=$\mathcal{L}^T_{rec}\left(x^T_H,\hat{x}^T_H\right)+\mathcal{L}^H_{rec}\left(x^S_H,\hat{x}^S_{H'}\right)$ is proposed to align style while maintaining pathological and structural features.

In summary, through the self-supervised domain adaptation strategy, we can decouple quality-related, target style-related and content-related embedding, and further recombine content-related and style-related embedding so as to unify the style of the patch-wise fundus image while removing low-quality interference factors. As a result, the proposed method enables the improvement of fundus image quality without any high-quality reference images.

\subsection{Jointly Optimizing}
The final self-supervised loss $\mathcal{L}_{s}$ is designed to:
\begin{equation}
\begin{aligned}
\mathcal{L}_{s}=\mathcal{L}_Q+\mathcal{L}_T+\mathcal{L}_C,
\end{aligned}
\end{equation}
\noindent where $\mathcal{L_C}=\mathbb{E}_{x^S_{H}}\left[\|\hat{x}^S_{H1}-\hat{x}^S_{H'}\|_1\right]$ to constrain $\hat{x}^S_{H1}$ and $\hat{x}^S_{H'}$ to be similar, which establishes an intrinsic link between the quality-related and style-related disentanglement. Then the overall objective function $\mathcal{L}$ of DASQE can be formulated as:
\begin{equation}
\begin{aligned}
\mathcal{L} = \mathcal{L}_{s}+\lambda_1\mathcal{L}^{Q_e}_{adv}+\lambda_2\mathcal{L}^{Q_I}_{adv},
\end{aligned}
\end{equation}
\noindent where $\lambda_1=\lambda_2=0.1$ are regularization parameters to balance the losses $\mathcal{L}_{s}$,$\mathcal{L}^{Q_e}_{adv}$, and $\mathcal{L}^{Q_I}_{adv}$.

\begin{figure*}[t]
\centerline{\includegraphics[width=1.8\columnwidth]{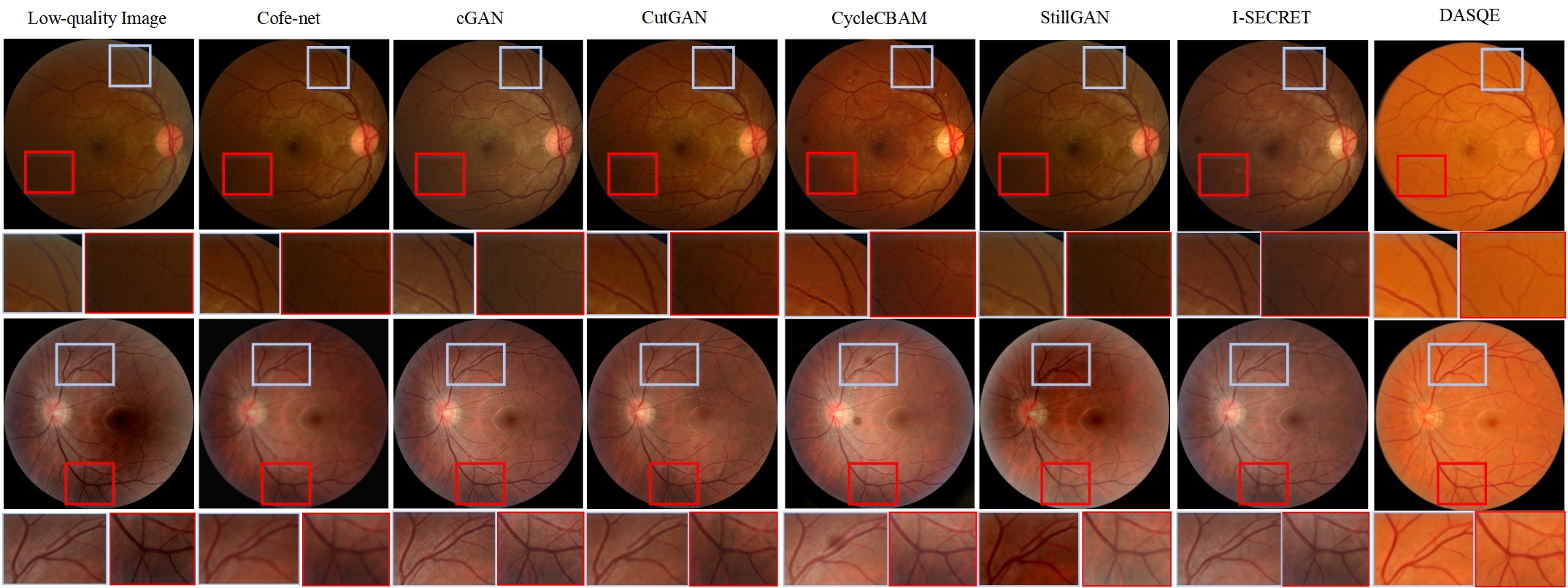}}
\caption{Visual comparisons on the low-quality image enhancement between the proposed and other deep learning methods. }
\label{contrast}
\end{figure*}

\section{Experiments}
In this section, we conduct experiments on two benchmarks including EyeQ \cite{fu2019evaluation} and Messidor datasets \cite{decenciere2014feedback} for evaluating the performance of quality enhancement of the proposed method. Moreover, we further investigate the effects of the enhanced images on extensive medical image analysis tasks, such as lesson segmentation, vessel segmentation, and disease grading.
\subsection{Datasets and Evaluation Metrics}
\noindent\textbf{EyeQ dataset} is a large-scale public benchmark for fundus image quality assessment, which consists of 28,792 fundus images with different quality labels, including high-quality, usable-quality and low-quality. It is worth mentioning that we re-label the images of "Usable-quality" with "low-quality" for pre-training the binary quality assessment network.\\ \noindent\textbf{Messidor dataset} includes 1,200 fundus images from three medical institutions, and the DR grading annotations are provided to measure the severity of diabetic retinopathy.\\
\noindent\textbf{Evaluation Metrics.} We use peak signal-to-noise ratio (PSNR) and structural index measure (SSIM) \cite{hore2010image} to evaluate the quality enhancement performance. In addition, it should be noted that for the quantitative metrics PSNR and SSIM, paired images are required. Due to the nonexistence of low-quality images in the relevant datasets, we follow the degradation pipeline proposed by \cite{shen2020modeling} to synthesize the corresponding low-quality dataset to measure the performance of image enhancement. In addition, in order to investigate the influence of higher quality images enhanced by our method on DR grading task, we also further train the disease grading network on the messdior dataset. The accuracy and quadratic weighted kappa are chosen to evaluate the grading performance.

\subsection{Implementation}
In this paper, we implement the proposed method using PyTorch with 4 NVIDIA Quadro RTX 6000 GPUs. The architecture of encoder and decoder is similar to the one in \cite{liu2017unsupervised}, and the weights are not shared across encoders. Due to the large-sized and diversity of the original images, all images are resized to 512 $\times$ 512, and the size of patch-wise fundus images is set to 128 $\times$ 128. During the training, we apply Adam optimizer to update the parameters with \emph{learning rate}=0.0001, \emph{momentum} = 0.9, \emph{batch size} = 32. The maximum number of training iterations is set to 300K. The weights of the proposed method are initialized from a Gaussian distribution $\mathcal{N}$(0, 0.02). 

\subsection{Comparison with State-of-the-Art Methods}

\begin{table}\tiny
    \centering
    \resizebox{\linewidth}{!}{
    \begin{tabular}{cccccc}
        \toprule
        Methods  & PSNR & SSIM & \makecell{Number of \\low-quality \\images} & \makecell{Number of \\reference \\images} & Types\\
        \midrule
        LIME(2016)     & 13.54 & 0.868 & / & /  &Traditional \\
        Fu et al.(2014) & 9.76 & 0.564 & / & /  &Traditional \\
        He et al.(2010) & 15.56 & 0.759 & / & / &Traditional  \\
        \midrule
        \midrule
        cofe-net(2020)   & 20.51 & 0.885  & 7196 & 7196 & Supervised\\
        cGAN(2017)       & 26.35 & 0.894  & 7196 & 7196 & Supervised\\
        I-SECRET(2021)   & 27.36 & 0.908  & 12543 & 7196 & Semi-Supervised\\
        CutGAN(2020)     & 22.76 & 0.872  & 7196 & 7196 & Unsupervised\\
        Cycle-CBAM(2019)   & 21.56 & 0.843  & 7196 & 7196 & Unsupervised\\
        StillGAN(2021)   & 25.38 & 0.894  & 7196 & 7196 & Unsupervised\\
        \midrule
        \midrule
        DASQE     & \textbf{28.47} & \textbf{0.914}  & 7196 & /& Unsupervised\\
        \bottomrule
    \end{tabular}
    }
    \caption{The comparison between our method with the SOTA methods for low-quality fundus image enhancement on EyeQ dataset.}
    \label{tab1}
\end{table}

In this part, we provide a comprehensive comparison of the proposed method with the traditional and deep learning-based methods. The traditional image enhancement methods include: LIME \cite{guo2016lime}, distribution fitting \cite{fu2014retinex}, and latent structure-drive \cite{he2010}. Deep learning-based approaches include: cofe-Net \cite{shen2020modeling}, I-SECRET \cite{cheng2021secret}, cGAN \cite{isola2017image}, CutGAN \cite{park2020contrastive}, CycleCAN \cite{you2019fundus}, and StillGAN \cite{ma2021structure}. Experimental results are reported in Table~\ref{tab1} where the best results are boldfaced.
The proposed method, DASQE, surpasses all the compared methods significantly, in terms of PSNR and SSIM without the guidance of any high-quality reference images. Although these comparable deep learning methods obtain better results than the traditional methods, the fact that they require pairwise high-quality images largely reduces their practicality for the problem that we aim to solve.
Notably, no high-quality images are available for our self-supervised learning method, and the supervision is derived from the image data itself, we also further visualize some cases of low-quality image enhancement for a range of comparable methods in Fig.\ref{contrast}. Although all methods exhibit decent ability of enhancing low-quality fundus images, unsupervised methods based on domain transfer suffer from apparent domain shifts. For example, the enhanced images obtained by Cycle-CBAM introduce undesired artifacts, which might mislead ophthalmologists to diagnose fundus disease. For StillGAN, the blurring of low-quality images is not well improved. In addition, due to the fact that the effect of style shifts on image enhancement is not considered, the comparable methods show limited improvement for uneven illumination. Our method presents remarkable performance on low-quality fundus image enhancement, and alleviates the domain shift problem, resulting in better visual perception. In order to prove the robustness of the proposed method for low-quality image enhancement on other datasets, we also report the experimental results of the proposed method on the Messidor dataset in Table~\ref{tab2}. From the results, we can observe that our method again achieves competitive performance against the other methods.

\begin{table}\tiny
    \centering
    \renewcommand{\arraystretch}{0.8}
    \resizebox{\linewidth}{!}{
    \begin{tabular}{cccccc}
        \toprule
        Methods  & PSNR & SSIM & Methods & PSNR & SSIM \\
        \midrule
        LIME     &12.36 & 0.852 & cGAN & 25.83  &0.867 \\
        Fu et al. &11.32 & 0.573 & CutGAN  & 24.37  &0.883 \\
        He et al. &16.23  & 0.764 & Cycle-CBAM  &23.56  &0.851 \\
        cofe-net  & 21.97  &0.852  & StillGAN   &26.32  &0.871  \\
        I-SECRET   &26.73  &0.874  & DASQE  & \textbf{29.36} & \textbf{0.897}   \\
        \bottomrule
    \end{tabular}
    }
    \caption{The comparison between our method with the SOTA methods for low-quality image enhancement on the Messidor dataset.}
    \label{tab2}
\end{table}

\begin{figure}[t]
\centerline{\includegraphics[width=0.9\columnwidth]{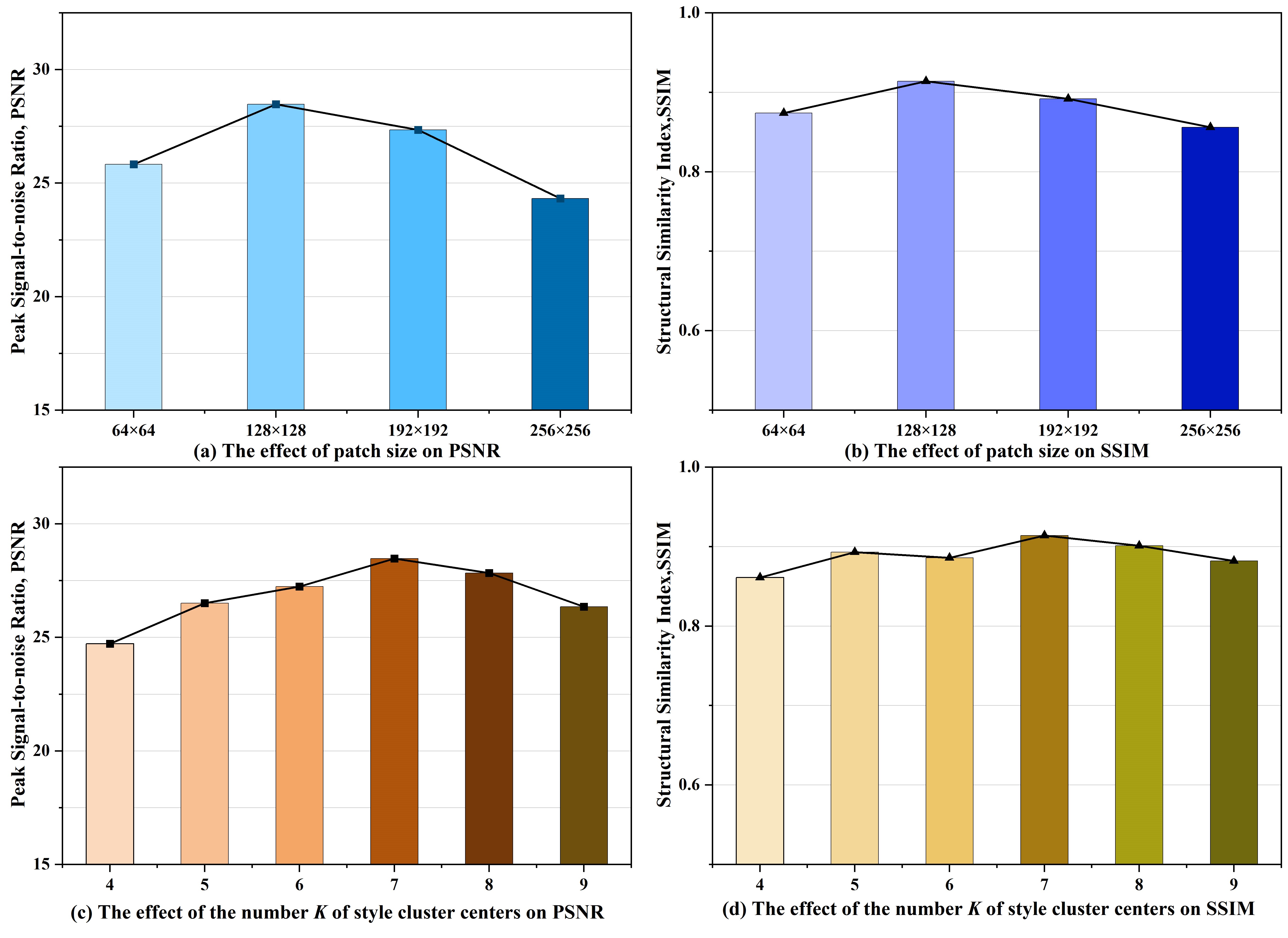}}
\caption{Comparison of low-quality fundus image enhancement results based on the different patch sizes (a-b) and different number \emph{K} of style cluster centers (c-d).}
\label{patchc}
\end{figure}

\subsection{Ablation Study}
In this section, we perform a series of ablation experiments to analyze the effect of patch-wise image domains and multiple feature decoupling for image quality enhancement. During the construction of the patch-wise image domain, there are two key hyper-parameters, including the size of the patch and the number of style cluster centers. We first explore the effect of the sizes of patches on quality enhancement. Based on the EyeQ dataset, we report the quality enhancement performance at different patch sizes in Fig.~\ref{patchc}(a-b). It can be found that the best performance metrics for quality enhancement are achieved when the patch size is 128$\times$128. Furthermore, both oversized and undersized patches result in degraded performance of image enhancement. The reason is that the number of low-quality regions is affected by the size of the patch-wise images. When the size of patches changes from 128$\times$128 to 64$\times$64, the number of high-quality patch-wise images increases. This may lead to more incorrect quality assessment results for pre-trained quality assessment networks. In contrast, when the size of patches changes from 128$\times$128 to 256$\times$256, the performance of quality enhancement degrades with the reduction in the number of high-quality patches.
 
In addition to the patch size, we also explore the intrinsic relationship between the number of style clusters and low-quality image enhancement as shown in Fig~\ref{patchc}(c-d). We observe that the proposed method achieves the best performance when the number of clusters is set to 7. More specifically, 1) when the number of the style clusters is set from [4, 6], the clustering happens to overlap, which leads to inconsistent styles of patch-level images in the target style domain. 2) when the number of style clusters exceeds 7, there are prominent outlier points in the clustering results, which increases the complexity of optimizing for style disentanglement.

\begin{table}\tiny
  \centering
  \renewcommand{\arraystretch}{0.8}
  \resizebox{\linewidth}{!}{
  \begin{tabular}{ccccc}
     \toprule
     \multirow{2}[1]{*}{Variants} & \multicolumn{2}{c}{EyeQ} & \multicolumn{2}{c}{Messidor} \\
     \cmidrule{2-5}          & \multicolumn{1}{c}{PSNR} & \multicolumn{1}{c}{SSIM} & \multicolumn{1}{c}{PSNR} & \multicolumn{1}{c}{SSIM} \\
     \midrule
     \makecell{Remove the quality-related\\disentanglement} & 16.32 & 0.768 & 17.63 & 0.758 \\
     \makecell{Remove the style-related \\disentanglement} & 19.24 & 0.832 & 18.52 & 0.817 \\
     DASQE  & 28.47 & 0.914 & 29.36      & 0.897 \\
     \bottomrule
  \end{tabular}%
  }
  \caption{Quantitative results for low-quality image enhancement.}
  \label{tab4}%
\end{table}%

\begin{figure}[t]
\centerline{\includegraphics[width=0.9\columnwidth]{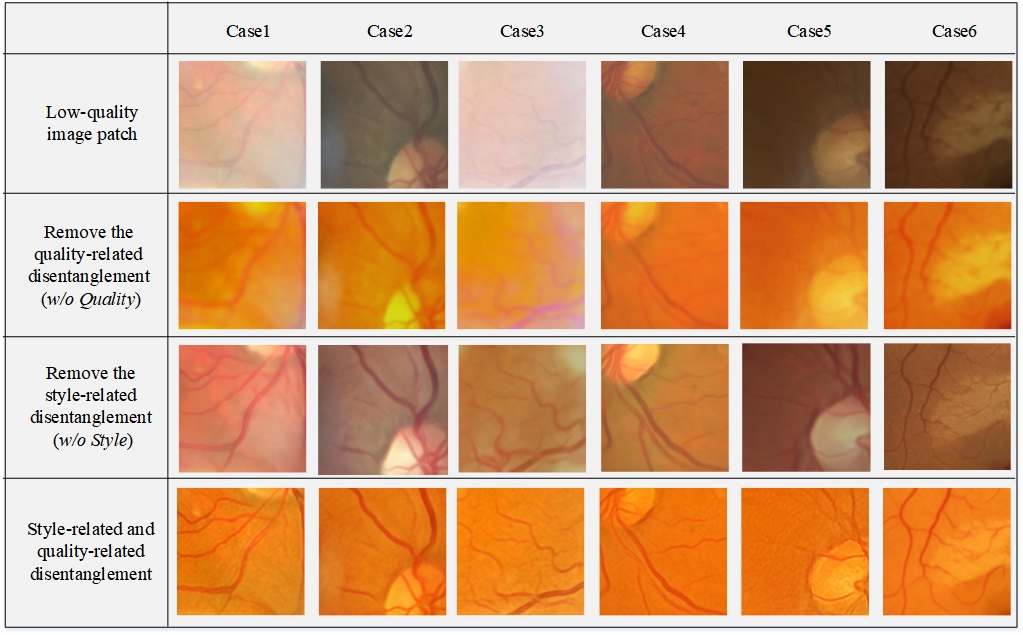}}
\caption{The effect of the different disentanglement branches for the patch-wise image quality enhancement.}
\label{patch}
\end{figure}

To investigate the effectiveness of multiple feature decoupling for image quality enhancement, we compare the proposed method with its two variants, respectively. Quantitative results are shown in Table \ref{tab4}. When the quality decoupling branch is removed, the performance metrics PSNR/SSIM for the quality enhancements decrease to 16.32/0.768 on the EyeQ dataset. This reveals that the method fails to effectively improve the quality of the fundus image. When we remove the style decoupling branch and keep the quality decoupling branch, the PSNR also drops to 19.24 on EyeQ dataset. Our results suggest that the simultaneously decoupling of both quality and style features is critical for improving the quality of the fundus image. To demonstrate the effectiveness of the different decoupling branches more intuitively, we also further visualize some patch-wise fundus images in Fig.\ref{patch}. 
As shown in the 2nd row in Fig.\ref{patch}, only the style-related disentanglement branch is incapable of enhancing the quality of the images. For example, the retinal vessels of patch-wise images do not appear clearer by introducing style disentanglement. The low-quality factors of images are also not eliminated. From the 3rd row in Fig.\ref{patch}, we can observe that although \emph{w/o Style} is able to transform general knowledge from patch-wise low-quality domains to patch-wise high-quality domains, there exist a large style gap for the enhanced patch-wise images, which is also detrimental to the quality enhancement of image-wise images.

\begin{table}\tiny
    \centering
    \renewcommand{\arraystretch}{0.8}
    \resizebox{\linewidth}{!}{
    \begin{tabular}{cccccc}
        \toprule
        Methods  & ACC & Kappa & Methods & ACC & Kappa \\
        \midrule
        \makecell{w/o enhancement} &0.681 & 0.657 & CutGAN  & 0.717      & 0.694 \\
        cofe-net & 0.704  & 0.687 & Cycle-CBAM   & 0.701 & 0.682 \\
        I-SECRET & 0.735  & 0.718  & StillGAN   & 0.724 & 0.706  \\
        cGAN & 0.731  & 0.712  & DASQE  & \textbf{0.752} & \textbf{0.734}   \\
        \bottomrule
    \end{tabular}
    }
    \caption{The DR grading performance on the Messidor dataset.}
    \label{tab5}
\end{table}

\subsection{Application}

\begin{figure}[H]
\centerline{\includegraphics[width=0.9\columnwidth]{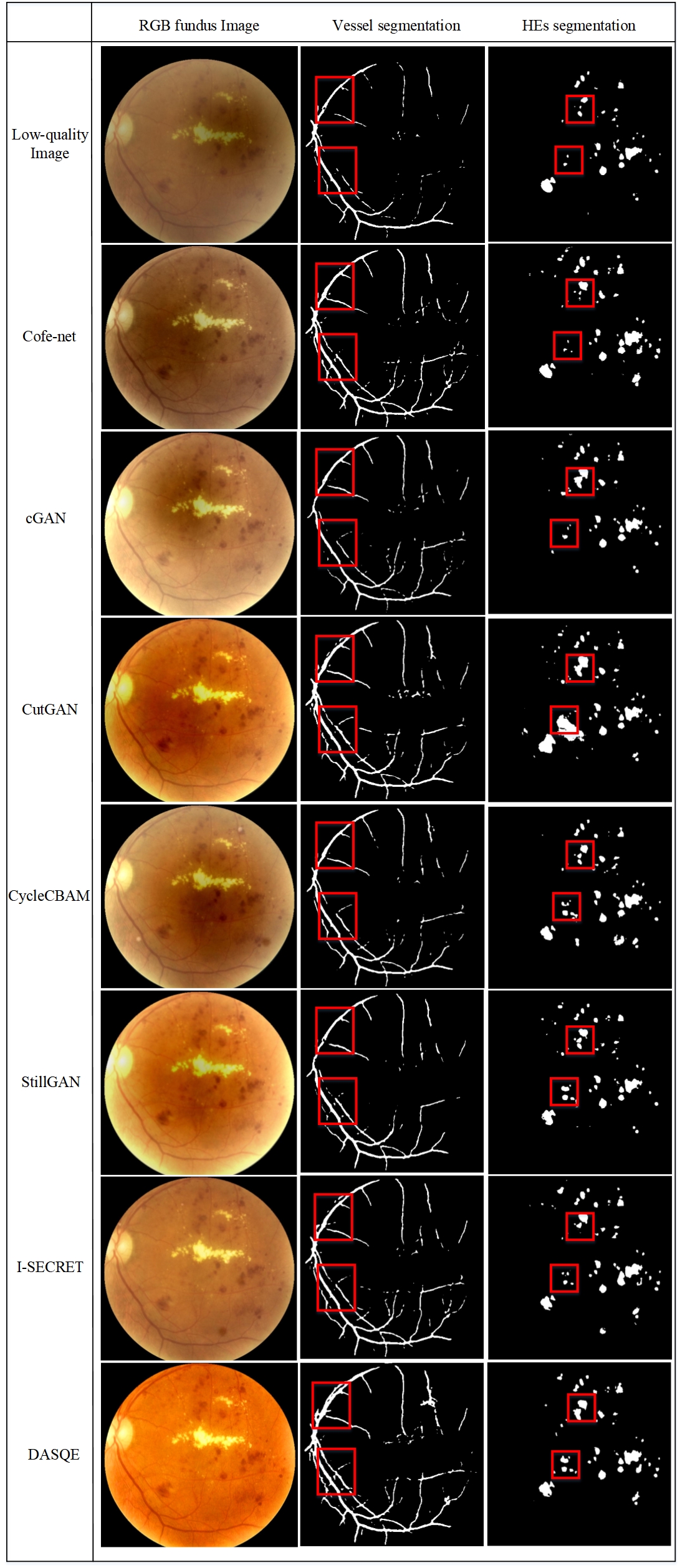}}
\caption{Visualization of Vessel and Lesion(HEs) segmentation on low-quality images and enhanced images.}
\label{apps}
\end{figure}

To investigate the effects of the DASQE on downstream medical image analysis tasks, we present a series of visual comparison results in Fig.~\ref{apps}. The experimental results show that the proposed method enables considerable performance improvements in the various fundus image analysis tasks, including vessel/lesion segmentation and disease grading.\\
\noindent\textbf{Vessel segmentation:} For computer-aided diagnosis of ophthalmic diseases, the precise segmentation of retinal vessels from fundus images is an important prerequisite. Such as, in \emph{proliferative diabetic retinopathy} (PDR), the number of neovascularisations is a key indicator to reflect its severity. For the vessel segmentation of fundus images, we test CE-Net \cite{gu2019net} pre-trained on DRIVE dataset \cite{staal2004ridge} for vessel segmentation on the enhanced images of proposed and comparable methods. As shown in the 2nd column of Fig.\ref{apps}, we visualize some cases of vessel segmentation. As can be seen that CE-Net can capture more fine-grained vessel segmentation results of the enhanced fundus images obtained by our proposed method DASQE.

\noindent\textbf{Lesion segmentation:} In \emph{non-proliferative diabetic retinopathy} (NPDR), the number of retinal hemorrhages (HEs) is an important indicator of disease severity. To investigate the effect of the proposed method on the segmentation of small lesions (HEs), we test UNet3+ \cite{huang2020unet} pre-trained on the IDRiD dataset \cite{porwal2018indian} for HEs segmentation on enhanced images. The segmentation results of HEs are shown in the 3rd column of Fig.\ref{apps}. Our method can produce enhanced images with more clear structures of clinical lesions, which can be precisely identified by ophthalmologists or automated diagnostic systems.  

\noindent\textbf{Disease grading:} To verify the influence of the quality improvement by our method on the disease grading, we further conduct the comparison among the improved-quality images  by competing methods using a DenseNet-121 \cite{huang2017densely} as the grading model. The DR grading results are shown in Table~\ref{tab5}. We observe that our method is the most beneficial for the disease grading of the classification model.
\section{Conclusion}
The enhancement methods of low-quality fundus images have long been considered to be a critical step before the diagnosis of eye disease. In this work, we propose a novel reference-free method to extend a self-supervised learning framework to the quality enhancement task. Our approach is the first to address the image quality enhancement problem from a complete self-supervised learning perspective. 
By combining the strengths of self-supervised learning and domain adaptation, the proposed approach not only significantly improves the state-of-the-art results in image quality enhancement on multiple benchmark datasets but also effectively aligns the illumination style of the fundus images. Herein, we focus on this more challenging and more widely applicable approach. The results show that our DASQE framework can produce high-quality images of better generalizability and robustness compared to state-of-the-art quality improvement methods on the various image analysis tasks. We strongly believe that the presented line of research is worth pursuing further. 

\section*{Acknowledgments}
This research was supported by the National Natural Science Foundation of China (No.62076059), and the Science Project of Liaoning province (2021-MS-105).
\bibliographystyle{named}
\bibliography{ijcai22}
\end{document}